\documentclass[twocolumn]{aastex63}
\usepackage{times}
\usepackage{amsmath}
\usepackage{textcomp}
\usepackage{amssymb}
\usepackage{natbib}
\usepackage{float}
\usepackage{color}
\usepackage{graphicx}
\newcommand{\Msun}{\ensuremath{M_{\odot}}}
\newcommand{\lum}{erg\,s$^{-1}$}

\newcommand{\nustar}{NuSTAR}
\newcommand{\xmm}{XMM-Newton}

\newcommand{\ergflux}{\mbox{${\rm \, erg \,\, cm^{-2} \, s^{-1}}$}}
\newcommand{\gm}{$\gamma$}
\newcommand{\hbeta}{H{$\beta$}}
\newcommand{\halpha}{H{$\alpha$}}

\newcommand{\NeIII}{[Ne{\sevenrm\,III}]\,$\lambda$3869}

\newcommand{\OIIIb}{[O{\sevenrm\,III}]\,$\lambda$5007}
\newcommand{\OIIIab}{[O{\sevenrm\,III}]\,$\lambda\lambda$4959,5007}
\newcommand{\OIIab}{[O{\sevenrm\,II}]\,$\lambda\lambda$3727,3729}

 \font\sevenrm=cmr7 scaled 1000

\received{\today}
\revised{\today}
\accepted{\today}
\submitjournal{ApJ}

\shorttitle{A Serendipitous NuSTAR detection of a GRS: J1128+5831}
\shortauthors{Paliya et al.}

\begin{document}
\title{A Serendipitous NuSTAR Detection of a Giant Radio Source Harboring an Obscured Active Galactic Nucleus}

\correspondingauthor{Vaidehi S. Paliya}
\email{vaidehi.s.paliya@gmail.com}

\author[0000-0001-7774-5308]{Vaidehi S. Paliya}
\affiliation{Inter-University Centre for Astronomy and Astrophysics (IUCAA), SPPU Campus, Pune 411007, India}

\author[0000-0001-5544-0749]{S. Marchesi}
\affiliation{Dipartimento di Fisica e Astronomia (DIFA), Università di Bologna, via Gobetti 93/2, I-40129 Bologna, Italy}
\affiliation{Clemson University, Clemson, SC 29634, USA}
\affiliation{INAF, Osservatorio di Astrofisica e Scienza dello Spazio di Bologna, via P. Gobetti 93/3, 40129 Bologna, Italy}

\author[0000-0002-7791-3671]{X. Zhao}
\affiliation{Cahill Center for Astrophysics, California Institute of Technology, 1216 East California Boulevard, Pasadena, CA 91125, USA}

\author[0000-0002-4464-8023]{D. J. Saikia}
\affiliation{Inter-University Centre for Astronomy and Astrophysics (IUCAA), SPPU Campus, Pune 411007, India}
\affiliation{ Fakult\"at f\"ur Physik, Universit\"at Bielefeld, Postfach 100131, 33501 Bielefeld, Germany}
\affiliation{Department of Physics, Assam Don Bosco University, Tapesia Campus, Guwahati 781017, Assam, India}

\author{Moumita Pal}
\affiliation{Department of Physics, Ashoka University, Sonepat, Haryana 131029, India}

\author[0000-0002-4864-4046]{Somak Raychaudhury}
\affiliation{Department of Physics, Ashoka University, Sonepat, Haryana 131029, India}
\affiliation{Inter-University Centre for Astronomy and Astrophysics (IUCAA), SPPU Campus, Pune 411007, India}
\affiliation{School of Physics and Astronomy, University of Birmingham, Birmingham B15 2TT, UK}

\begin{abstract}
Giant radio sources (GRSs) harbor the Universe’s largest structures generated by individual galaxies, with projected source sizes exceeding 700 kpc. These enigmatic objects have been mainly studied at radio frequencies, and their physical properties in the high-energy domain are poorly understood. Here we present the results of a multiwavelength study focused on NuSTAR J112829+5831.8 (J1128+5831), the only known GRS serendipitously detected with the Nuclear Spectroscopic Telescope Array. Being located in proximity to the famous interacting galaxy system, Arp 299, J1128+5831 has been serendipitously observed also by the Chandra X-ray Observatory, Hubble Space Telescope, and XMM-Newton satellites. From radio observations with the Low Frequency Array, the NRAO VLA Sky Survey and the Very Large Array Sky Survey, we have determined that J1128+5831 has an overall steep radio spectrum ($\alpha=-0.86$; $F_\nu\propto\nu^\alpha$) and a low core dominance ($C_{\rm D}=-2.4$, in log-scale), indicating the source to be viewed at large angles. From the X-ray spectral analysis, we found J1128+5831 to harbor an obscured active galactic nucleus (AGN) with neutral hydrogen column density exceeding $10^{23}$ cm$^{-2}$. Its optical spectrum, taken with the Dark Energy Spectroscopic Instrument, exhibits prominent narrow emission lines but lacks broad components, thus confirming J1128+5831 to be a Type 2 AGN powered by a radiatively efficient accreting system. Overall, the broadband properties of J1128+5831 are consistent with those observed for the general GRS population.
\end{abstract}

\section{Introduction}
Active Galactic Nuclei (AGNs) are among the most intriguing astrophysical objects, which hold the key to understanding the growth and evolution of supermassive black holes and galaxies harboring them \citep[see, e.g.,][]{2012ARA&A..50..455F,2013ARA&A..51..511K}. About 10\% of AGNs host powerful relativistic jets that are best resolved at radio wavelengths. The objects hosting closely aligned jets (jet viewing angle $\theta_{\rm v}\lesssim1/\Gamma$, where $\Gamma$ is the bulk Lorentz factor) are termed as blazars, and those observed at large angles are often called radio galaxies or quasars. Other than in the radio band, extended jets have been detected at optical, X-ray, and also at \gm-ray energies \citep[e.g.,][]{2000ApJ...540L..69S,2010Sci...328..725A,2018ApJ...856...66M,2024ApJ...965..163Y}.

Based on the radio power and morphology, radio sources have been divided into two classes: Fanaroff-Riley (FR) type I and II \citep[][]{1974MNRAS.167P..31F}. The former are low-luminosity AGNs, often associated with radiatively inefficient accretion systems, and exhibit brighter radio emission nearer to the cores and less well-collimated radio jets which expand to form diffuse lobes of emission. On the other hand, powerful FR~II objects often show bright, edge-brightened lobes on opposite sides, well-collimated jets with overall sizes extending up to Megaparsec scales, and are usually powered by a radiatively efficient accretion process. However, sensitive and high-resolution low-frequency observations taken with the Low Frequency Array (LOFAR) have revealed a complex pattern with no apparent connection between radio morphologies and accretion state and radio power \citep[][]{2019MNRAS.488.2701M,2022MNRAS.511.3250M,2025MNRAS.539..463C}.

The relativistic jets in some of the radio sources lead to megaparsec (Mpc)-scale structures. Such objects, with an overall projected source size exceeding 0.7 Mpc, are termed giant radio sources (GRSs), including galaxies and quasars \citep[][]{1974Natur.250..625W}. These enigmatic objects often exhibit FR~II radio morphology and are usually found to have low levels of accretion activity, though exceptions are known \citep[e.g.,][]{1999MNRAS.309..100I,2012MNRAS.426..851K,2020A&A...642A.153D}. These objects typically reside in elliptical galaxies harboring massive black holes ($\gtrsim10^8$ \Msun); however, a handful of them are also found to be associated with spiral galaxies \citep[e.g.,][]{2011MNRAS.417L..36H,2014ApJ...788..174B,2025A&A...699L...4S}.

Several models have been put forward to explain the formation of such gigantic radio structures. For example, GRSs might be either residing in sparse environments enabling unimpeded growth of jets \citep[see, e.g.,][]{2015MNRAS.449..955M} or have powerful jets \citep[cf.][for a review]{2023JApA...44...13D}. The presence of Mpc-scale jets can also be due to episodic AGN activity \citep[cf.][]{2000MNRAS.315..371S,2025A&A...696A..97D}. Furthermore, with the advent of sensitive radio telescopes, e.g., LOFAR and Rapid ASKAP Continuum Survey, the number of known GRSs has considerably increased in the last decade, and more than 11000 radio sources have been reported to host Mpc-scale structures \citep[][]{2024A&A...691A.185M}.

Compared to the radio band, the properties of GRSs at high-energies, namely X- and \gm-ray energies are less understood. The detection of the extended X-ray emission from radio lobes/jets of GRSs, likely due to inverse-Compton scattering of the cosmic microwave background photons, has been reported \citep[e.g.,][]{2008MNRAS.386.1774E,2015MNRAS.453.2438T,2023A&A...672A.179B}. \citet[][]{2018MNRAS.481.4250U} studied 14 GRSs selected from the INTEGRAL hard X-ray survey and found their core X-ray emission to be produced from a Comptonizing corona irradiated by a radiatively efficient accretion disk. Furthermore, \citet[][]{2025ApJ...989...36P} carried out a systematic search of the \gm-ray emitting GRSs and reported the identification of 16 objects. These objects expected to have a smaller $\theta_{\rm v}$ compared to non-\gm-ray detected GRS population, since the \gm-ray emission is highly sensitive to $\theta_{\rm v}$ \citep[][]{2015ApJ...810L...9L,2017ApJ...851...33P}.

To further explore the X-ray properties of GRSs, we utilized the largest catalog of GRSs published by \citet[][]{2024A&A...691A.185M} which contains 11585 sources identified using the second data release of the LOFAR Two Metre Sky Survey (LOTSS-DR2) and those reported in earlier works. We cross-matched the GRS catalog with the recently released catalog of 1274 hard X-ray sources detected serendipitously in the first 80 months of observations by the Nuclear Spectroscopic Telescope Array \citep[NuSTAR;][]{2024ApJS..273...20G}. The primary goal of this exercise was to identify any GRS serendipitously detected with NuSTAR. About 17\% serendipitously NuSTAR detected objects lie in the sky area covered by the LOTSS-DR2. We chose a search radius of 5 arcseconds and considered infrared (IR) counterpart coordinates provided in the 80-month NuSTAR serendipitous catalog (NSS80) and host galaxy coordinates given in the LOFAR GRS catalog. Only one object, NuSTAR J112829+5831.8 (hereafter J1128+5831, right ascension = 172$^{\circ}$.12298, declination = 58$^{\circ}$.5341, J2000), was found which was detected in all three NuSTAR bands of 3$-$8 keV, 8$-$24 keV, and 3$-$24 keV\footnote{The cross-matching of the NSS80 and LOFAR GRS catalogs led to the identification of another source, NuSTAR J153638+5750.2. However, it was undetected in 3$-$8 keV and 3$-$24 keV bands, marginally detected in the 8$-$24 keV band, and has no reported soft X-ray counterpart, making its hard X-ray detection questionable.}. The same result was also obtained by using the CDS xMatch service \citep[][]{2020ASPC..522..125P}. Therefore, J1128+5831 is likely to be the only GRS serendipitously detected with NuSTAR.

Given the paucity of high-energy observations of GRSs, we collected and analyzed all of the available multi-frequency observations of J1128+5831, and here we present the first detailed broadband study of this enigmatic GRS. Throughout, we adopted the flux density $F_{\nu}\propto\nu^{\alpha}$, where $\alpha$ is the radio spectral index. A flat cosmology with $H_0 = 70~{\rm km~s^{-1}~Mpc^{-1}}$ and $\Omega_{\rm M} = 0.3$ was used.

\section{NuSTAR J112829+5831.8}\label{sec2}
J1128+5831 lies 1.6 arcminutes south of the famous interacting galaxy system Arp 299 \citep[][]{1996ARA&A..34..749S}. This luminous IR galaxy system was the subject of extensive multiwavelength campaigns including observations taken with the Chandra X-ray observatory, Hubble space telescope (HST), ground-based radio telescopes, and recently NuSTAR \citep[][]{2003ApJ...594L..31Z,2010A&A...519L...5P,2013ApJ...779L..14A,2015ApJ...800..104P}. Some of these datasets covered the location of J1128+5831, thus leading to its serendipitous observation. In particular, \citet[][]{2015ApJ...800..104P} reported the serendipitous detection of J1128+5831 with NuSTAR. \citet[][]{2017ApJ...836...99L} carried out optical spectroscopic observation of J1128+5831 as part of the 40-month NuSTAR Serendipitous catalog and measured a redshift of $z=0.4103$. This object is also included in the fourteenth data release of the fourth \xmm~serendipitous source catalog and second Chandra source catalog as 4XMM~J112829.5+583202 and 2CXO~J112829.5+583202, respectively \citep[][]{2020A&A...641A.136W,2024ApJS..274...22E}. 

\begin{figure}
\includegraphics[width=\linewidth]{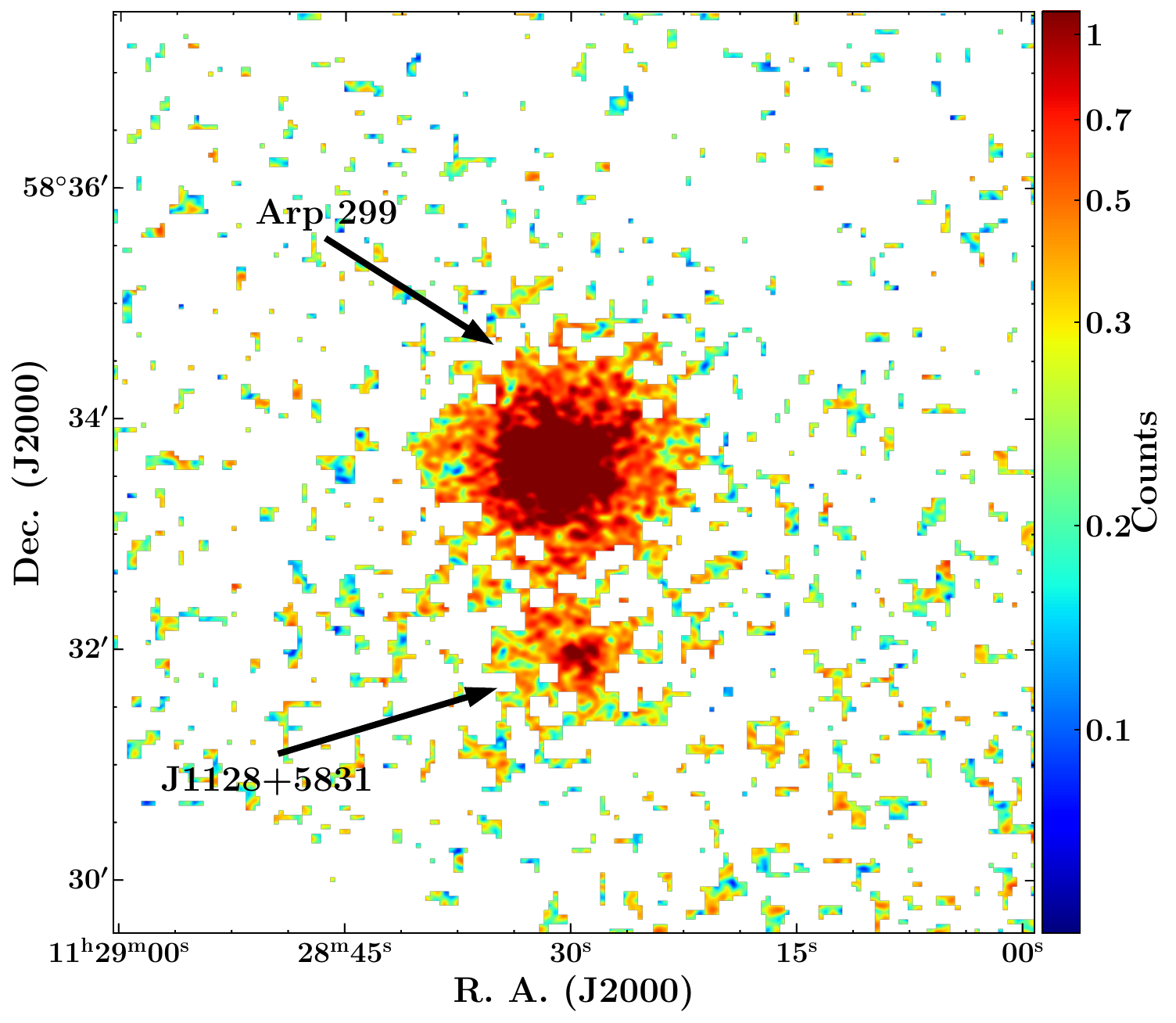}
\caption{3$-$79 keV NuSTAR image of Arp 299 region, generated by combining both focal plane module datasets. Color scaling is adjusted to highlight J1128+5831.}\label{fig:nustar_map}
\end{figure}

J1128+5831 is present in the LOFAR GRS catalog published by \citet[][]{2024A&A...691A.185M}. They reported the largest angular size and the projected jet length to be 2.64 arcminutes and $\sim$808 kpc, respectively, using the photometric redshift of $z=0.351$ \citep[][]{2023A&A...678A.151H}. On the other hand, using the spectroscopic redshift of $z=0.4103$, the derived projected jet length is 864 kpc.

\begin{table*}
\centering
\caption{Log of X-ray Observations. \label{tab:data}}
\begin{tabular}{lcccc}
\hline
\hline
Telescope & Sequence & Start Time & End Time & Exposure\\
 & ObsID & (UTC) & (UTC) & (ksec)\\
\hline
NuSTAR   & 50002041002  & 2013-03-12T23:21:07 & 2013-03-13T04:26:07 & 9 \\
         & 50002041003  & 2013-03-13T04:26:07 & 2013-03-14T13:01:07 & 60 \\
\xmm  & 0112810101 & 2001-05-06T23:26:20 & 2001-05-07T05:26:39 & 22 \\
      & 0679381101 & 2011-12-15T02:16:41 & 2011-12-15T04:59:12 & 12 \\
      & 0861250101 & 2020-05-08T21:01:52 & 2020-05-09T10:47:53 & 53 \\
      & 0861250201 & 2020-05-22T19:41:02 & 2020-05-23T08:54:50 & 49 \\
      & 0861250301 & 2020-11-22T12:40:11 & 2020-11-23T00:30:19 & 44 \\
Chandra  & 1641  & 2001-07-13T11:07:37 & 2001-07-13T18:29:55 & 25 \\
         & 6227  & 2005-02-14T04:19:26 & 2005-02-14T07:58:49 & 10 \\
         & 15619 & 2013-03-12T23:13:03 & 2013-03-13T10:23:16 & 39 \\
         & 15077 & 2013-03-13T21:56:06 & 2013-03-14T13:07:44 & 53 \\
\hline
\end{tabular}
\end{table*}

\section{Data Reduction and Collection}\label{sec3}
\subsection{NuSTAR}
NuSTAR observed the Arp 299 region on 2013 March 12 and 13 for $\sim$9 and $\sim$60 ksec, respectively (Table~\ref{tab:data}). To generate the image shown in Figure~\ref{fig:nustar_map}, we reduced both datasets; whereas, the second observation with longer exposure was considered to extract products needed for spectral analysis. The tool {\tt nupipeline}, built in NUSTARDAS, was used to reduce the raw NuSTAR data and for cleaning and calibrating the event files for both focal plane modules, FPMA and FPMB. To run the pipeline {\tt nuproducts}, we considered source and background regions of 30 and 70 arcseconds radii, respectively, from the same chip. FPMA and FPMB spectra were binned to have at least 15 counts per bin using the tool {\tt grppha}. We used XSPEC for spectral fitting and estimated the uncertainties at the 90\% confidence level. J1128+5831 was detected up to 20 keV in the NuSTAR data.

\subsection{XMM-Newton}
The Arp 299 region was observed with \xmm~five times, thrice in 2020, and once each in 2001 and 2011 (Table~\ref{tab:data}). We followed the standard procedures to analyze the EPIC-PN data using \xmm~Science Analysis Software (21.0.0). The pipeline {\tt epproc} and {\tt evselect} were used to obtain the calibrated and concatenated event list and remove the time periods of high background flares, respectively. We chose circular regions with a radius of 40 arcseconds to extract the source and background spectra. In particular, the background region was selected from the same chip but free from source contamination. We used the tasks {\tt rmfgen} and {\tt arfgen} to generate the redistribution matrix and ancillary response files. Similar steps were followed to reduce the EPIC-MOS data. Finally, we combined the individual EPIC-PN spectra using the task {\tt epicspeccombine} and the same command was used to combine MOS spectra separately. The stacked spectra were rebinned using {\tt grppha} with 15 counts per bin. The source was detected in the full 0.3$-$10 keV energy band in the \xmm~data.

\subsection{Chandra}
Chandra observed the Arp 299 region twice in 2013 (simultaneous with NuSTAR) and once in 2001 and 2005 each (Table~\ref{tab:data}). We analyzed the data acquired with the Advanced CCD Imaging Spectrometer (ACIS, 0.5–7 keV) using the Chandra Interactive Analysis of Observations (4.17). The tool {\tt chandra\_repro} was adopted to clean and calibrate the event files and we used the command {\tt specextract} to generate the source and background spectra. For the latter, the source and background regions were selected as circular regions of 3 arcseconds and 10 arcseconds radii, respectively. We also combined all ACIS-S observations using the task {\tt merge\_obs} to generate the 0.5$-$7 keV exposure-corrected image of the Arp 299 region. Individual spectra were stacked using the tool {\tt combine\_spectra}. The extracted spectrum was rebinned with 15 counts per bin, taking into account the background, using {\tt grppha}. J1128+5831 was detected above 1\,keV in the Chandra data.
\subsection{Optical}
We collected the HST image of Arp 299 region from the Mikulski Archive for Space Telescopes portal\footnote{\url{https://mast.stsci.edu/portal/Mashup/Clients/Mast/Portal.html}}. The observation was taken on 2019 April 28 for a net exposure of 2360 seconds using Advanced Camera for Surveys in F814W filter (central wavelength = 8045 \AA).

The optical spectrum of J1128+5831 was recently published as part of the Dark Energy Spectroscopic Instrument (DESI) data release 1 \citep[][]{2025arXiv250314745D}. The source was observed with the Mayall 4-meter telescope at Kitt Peak National Observatory on 2022 April 14 for a total exposure of 720 seconds. Further details about the DESI setup can be found on the survey website\footnote{\url{https://data.desi.lbl.gov/doc/releases/dr1/}}. We downloaded the optical spectrum from the DESI archive and used it to explore the central engine properties, e.g., black hole mass and accretion rate, of J1128+5831.

\subsection{Radio}
J1128+5831 lies in the footprint of several radio surveys. We downloaded the 10$\times$10 arcmin$^{2}$-sized LOFAR cutout image of the target from the LOFAR data release 2 website\footnote{\url{https://lofar-surveys.org/dr2_release.html}}. LOFAR operates in the 120$-$168 MHz frequency band (central frequency = 144 MHz), achieving a resolution of 6 arcseconds and a median rms sensitivity of 83 $\mu$Jy beam$^{-1}$ \citep[][]{2022A&A...659A...1S}.

We obtained a 10$\times$10 arcmin$^{2}$-sized cutout image from the ongoing Very Large Array Sky Survey\footnote{\url{http://cutouts.cirada.ca/}} (VLASS). This survey operates in the 2-4 GHz frequency band (central frequency = 3 GHz) and observes the sky north of $-$40$^{\circ}$ declination. The survey has achieved an angular resolution of $\sim$2.5 arcseconds, and the typical rms sensitivity for a single epoch is 120 $\mu$Jy beam$^{-1}$ \citep[][]{2020PASP..132c5001L}. Since VLASS has observed the same region of the sky multiple times to study radio transients, we downloaded all three epochs of the data and combined individual frames to generate a stacked image of J1128+5831.

We also extracted a 10$\times$10 arcmin$^{2}$-sized cutout image from the NRAO VLA Sky Survey (NVSS) database\footnote{\url{http://skyview.gsfc.nasa.gov/current/cgi/query.pl}}. This low-resolution (beam size 45 arcseconds) survey was conducted at 1.4 GHz covering the sky north of $-$40$^{\circ}$ declination \citep[][]{1998AJ....115.1693C}.

\section{Results and Discussion}\label{sec4}
J1128+5831 was serendipitously detected with NuSTAR when the satellite was observing Arp 299. We show the count map of the region by combining both focal plane modules datasets in Figure~\ref{fig:nustar_map}. Below, we describe the results obtained from the multiwavelength data analysis of the source.

\subsection{A Giant Radio Source}
\begin{figure}
\includegraphics[scale=0.4]{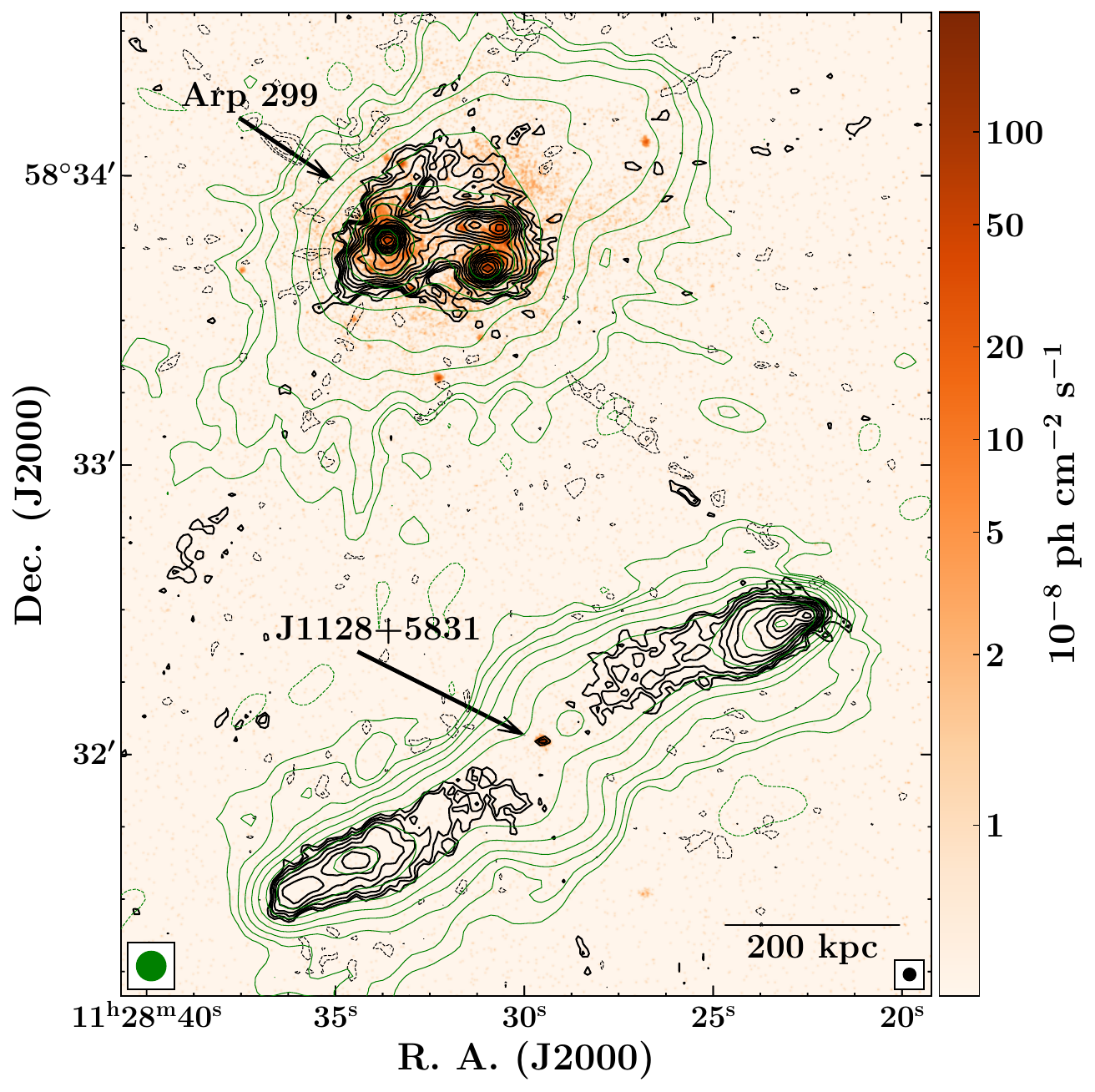}
\caption{The LOFAR (green) and VLASS (black) contours overplotted on the 0.5$-$7 keV image of the Arp 299 region using the data taken from the Chandra X-ray observatory. Contour levels are drawn at 3$\times$ rms $\times$ ($-$2,$-$1, 1) and increase in multiples of 2 for LOFAR and in multiples of $\sqrt{2}$ for VLASS data. The X-ray data are in native 0.492 arcseconds per pixel and smoothed using bi-cubic interpolation. The restoring beams of LOFAR and VLASS are shown at the bottom corners.}\label{fig:maps}
\end{figure}

In Figure~\ref{fig:maps}, we show the 0.5$-$7 keV image of the Arp 299 region taken with the Chandra satellite and overplot LOFAR and VLASS radio contours. A bipolar radio structure of J1128+5831 is evident. A bright radio hotspot and lobe were identified along the northwest direction in VLASS and LOFAR datasets. In contrast, the extended emission appears somewhat diffuse with no prominent hotspot in the southeast direction. Furthermore, X-ray emission was detected only from the core region.

A weak radio core would be difficult to isolate in the presence of strong bridge emission in the LOFAR and NVSS observations. It was also not detected in individual VLASS maps that have a local rms sensitivity of $\sim$0.15 mJy beam$^{-1}$. However, the core was detected in the stacked VLASS image with a flux density of $0.46\pm0.15$ mJy. We estimated the rest-frame core dominance at 3 GHz using the core flux density measured by VLASS and the total flux density computed from the NVSS map. In particular, we used the following equation \citep[see, e.g.,][]{2024ApJ...976..120P}:

\begin{equation}
    C_{\rm D}=\log\left(\frac{F_{\rm core}}{F_{\rm ext}}(1+z)^{\alpha_{\rm core}-\alpha_{\rm ext}} \right),
\end{equation}\label{eq:1}

The spectral indices of the core and extended emissions, $\alpha_{\rm core}$ and $\alpha_{\rm ext}$, respectively, were adopted to be $\alpha_{\rm core}=0$ and $\alpha_{\rm ext}=-0.8$. In Equation~\ref{eq:1}, the core and extended flux densities are quoted with parameters $F_{\rm core}$ and $F_{\rm ext}$ ($=F_{\rm total}-F_{\rm core}$), respectively, and were derived at rest-frame of 3 GHz. This exercise resulted in a core dominance of $C_{\rm D}=-2.4$. Such an extremely low value of core dominance indicates a large viewing angle of the jet since the lobe emission is usually isotropic, whereas the core emission is sensitive to the beaming effect \citep[cf.][]{1997MNRAS.284..541M}. Indeed, an anti-correlation of the source size and core dominance has been reported in the literature, suggesting that GRSs are typically viewed at larger angles \citep[e.g.,][]{1982JApA....3..465K,2022A&A...660A..59M,2023JApA...44...13D}.

We calculated the overall radio spectral index of J1128+5831 using total flux densities measured at 1.4 GHz (NVSS) and 144 MHz (LOFAR) and found it to be $\alpha=-0.86$. Such a steep radio spectrum is typically observed from GRSs \citep[][]{2020A&A...642A.153D}. Furthermore, the $k$-corrected 144 MHz radio power was derived from the following formula
\begin{equation}\label{eq:power}
 P_{\rm 144~MHz}=4\pi d^2_{\rm L}F_{\rm 144~MHz}(1+z)^{-(1+\alpha)},
\end{equation}
 where $F_{\rm 144~MHz}$ is the flux density measured by LOFAR survey and $d_{\rm L}$ is the luminosity distance. We estimated $P_{\rm 144~MHz}$ to be $1\times10^{27}$ W Hz$^{-1}$, which implies that J1128+5831 belong to the  powerful FR~II class. 

 The observation of a low core dominance and steep radio spectrum suggests J1128+5831 to be viewed at a large angle from the jet axis, similar to Type 2 AGNs. Such objects are usually obscured in nature, i.e., with a large neutral hydrogen column density. We further explored this possibility below. 
\begin{deluxetable}{lcc}
\tabletypesize{\footnotesize}
\tablecaption{Summary of the Parameters Obtained from the Joint X-Ray Spectral Fitting. \label{tab:spec_param}}
\tablewidth{0pt}
\tablehead{
\colhead{Model} & 
\colhead{Chandra + NuSTAR} & 
\colhead{\xmm~+ NuSTAR} 
}
\startdata
$\chi^2$/dof & 91.21/98 & 612.16/612 \\
$C_{\rm inst.}$\tablenotemark{a} & 1.09$^{+0.23}_{-0.18}$ & 0.91$^{+0.05}_{-0.04}$ \\
$\Gamma$\tablenotemark{b} & 1.44$^{+0.23}_{-0.23}$ & 1.78$^{+0.15}_{-0.14}$ \\
$N_{\rm H}$\tablenotemark{c} & 12.02$^{+2.20}_{-1.94}$ & 15.58$^{+2.46}_{-2.19}$ \\
norm\tablenotemark{d} & 6.87$^{+5.51}_{-3.10}$ & 18.32$^{+7.62}_{-5.09}$ \\
$\zeta_{\rm scat}$\tablenotemark{e} & 2.13$^{+0.01}_{-0.01}$ & 5.72$^{+0.02}_{-0.02}$ \\
$kT$\tablenotemark{f} & -- & 0.77$^{+0.11}_{-0.09}$ \\
abundance\tablenotemark{g} & -- & 0.12$^{+0.12}_{-0.07}$ \\
line energy\tablenotemark{h}  & -- & 1.30$^{+0.03}_{-0.03}$ \\
line width\tablenotemark{i}  & -- & 0.03 \\
line norm\tablenotemark{j}  & -- & 1.09$^{+0.33}_{-0.32}$ \\
$F_{\rm 2-10~keV}$\tablenotemark{k} & 2.04$^{+0.16}_{-0.71}$ & 2.05$^{+0.21}_{-0.31}$ \\
$F_{\rm 10-20~keV}$\tablenotemark{l} & 2.22$^{+0.24}_{-0.66}$ & 2.24$^{+0.22}_{-0.33}$ \\
$L_{\rm 2-10~keV}$\tablenotemark{m}  & 7.91$^{+0.69}_{-1.98}$ & 8.07$^{+0.83}_{-1.24}$ \\
$L_{\rm 10-20~keV}$\tablenotemark{n}  & 10.74$^{+0.89}_{-2.28}$ & 10.76$^{+0.81}_{-1.27}$\\
\enddata
\tablenotetext{a}{Cross-calibration constant between NuSTAR and \xmm/Chandra.}
\tablenotetext{b}{Power-law photon index.}
\tablenotetext{c}{Intrinsic line of sight 	equivalent hydrogen column density, in 10$^{22}$ atoms cm$^{-2}$.}
\tablenotetext{d}{Normalization of different model components at 1 keV, in 10$^{-5}$ ph cm$^{-2}$ s$^{-1}$ keV$^{-1}$.}
\tablenotetext{e}{Fraction of the AGN emission scattered by the obscuring medium, in per cent.}
\tablenotetext{f}{Temperature of the thermal component in the {\it mekal} model, in keV.}
\tablenotetext{g}{Relative abundance in the {\it mekal} model.}
\tablenotetext{h}{Central energy of the Gaussian line at rest-frame, in keV.}
\tablenotetext{i}{Width of the Gaussian line, in keV (fixed during the fitting).}
\tablenotetext{j}{Normalization of the Gaussian line, in 10$^{-6}$ ph cm$^{-2}$ s$^{-1}$.}
\tablenotetext{k}{2$-$10 keV energy flux, in 10$^{-13}$ \ergflux.}
\tablenotetext{l}{10$-$20 keV energy flux, in 10$^{-13}$ \ergflux.}
\tablenotetext{m}{Intrinsic 2$-$10 keV luminosity, in 10$^{43}$ \lum.}
\tablenotetext{n}{Intrinsic 10$-$20 keV luminosity, in 10$^{43}$ \lum.}
\end{deluxetable}

\begin{figure*}
\centering
\includegraphics[scale=0.35]{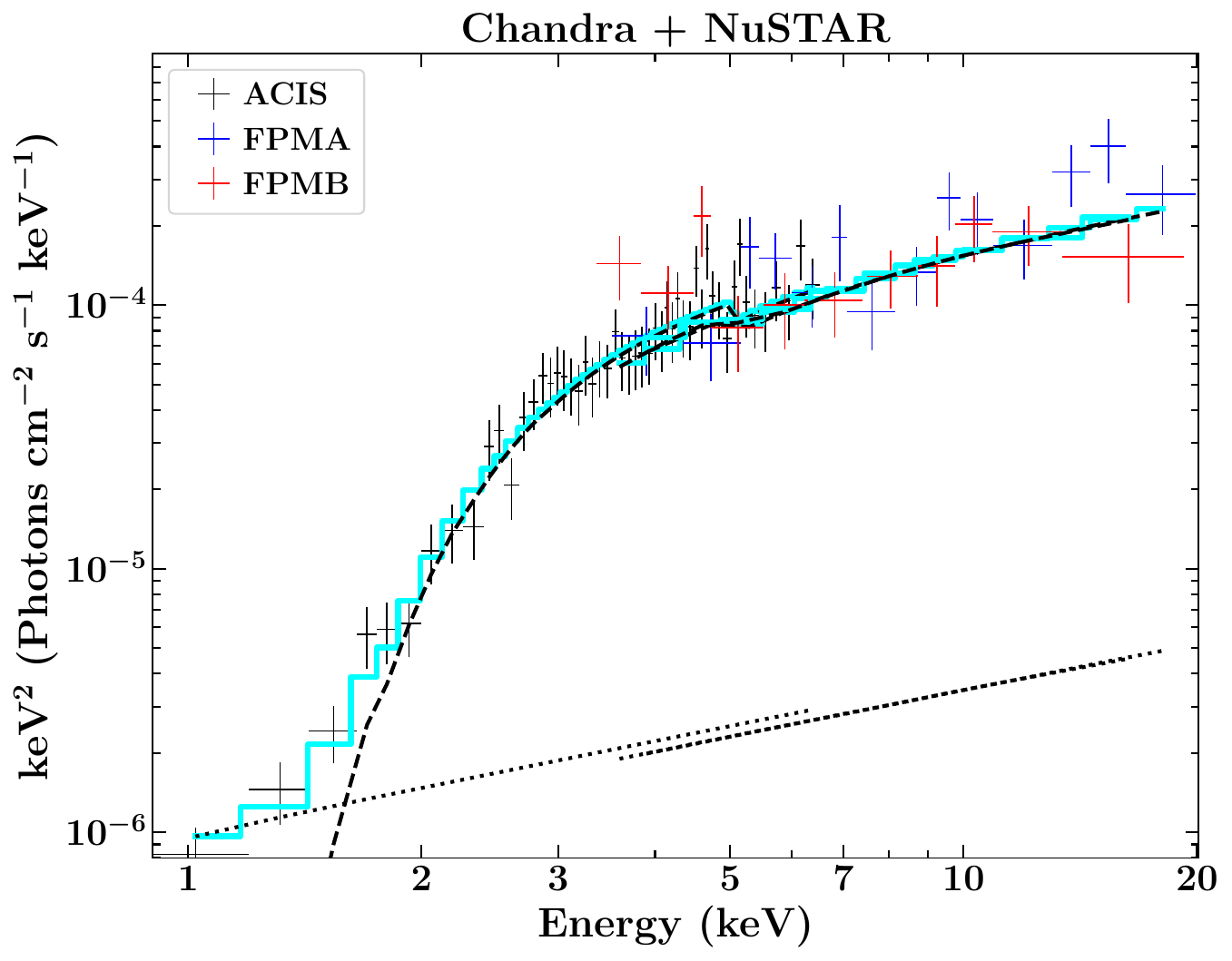}
\includegraphics[scale=0.35]{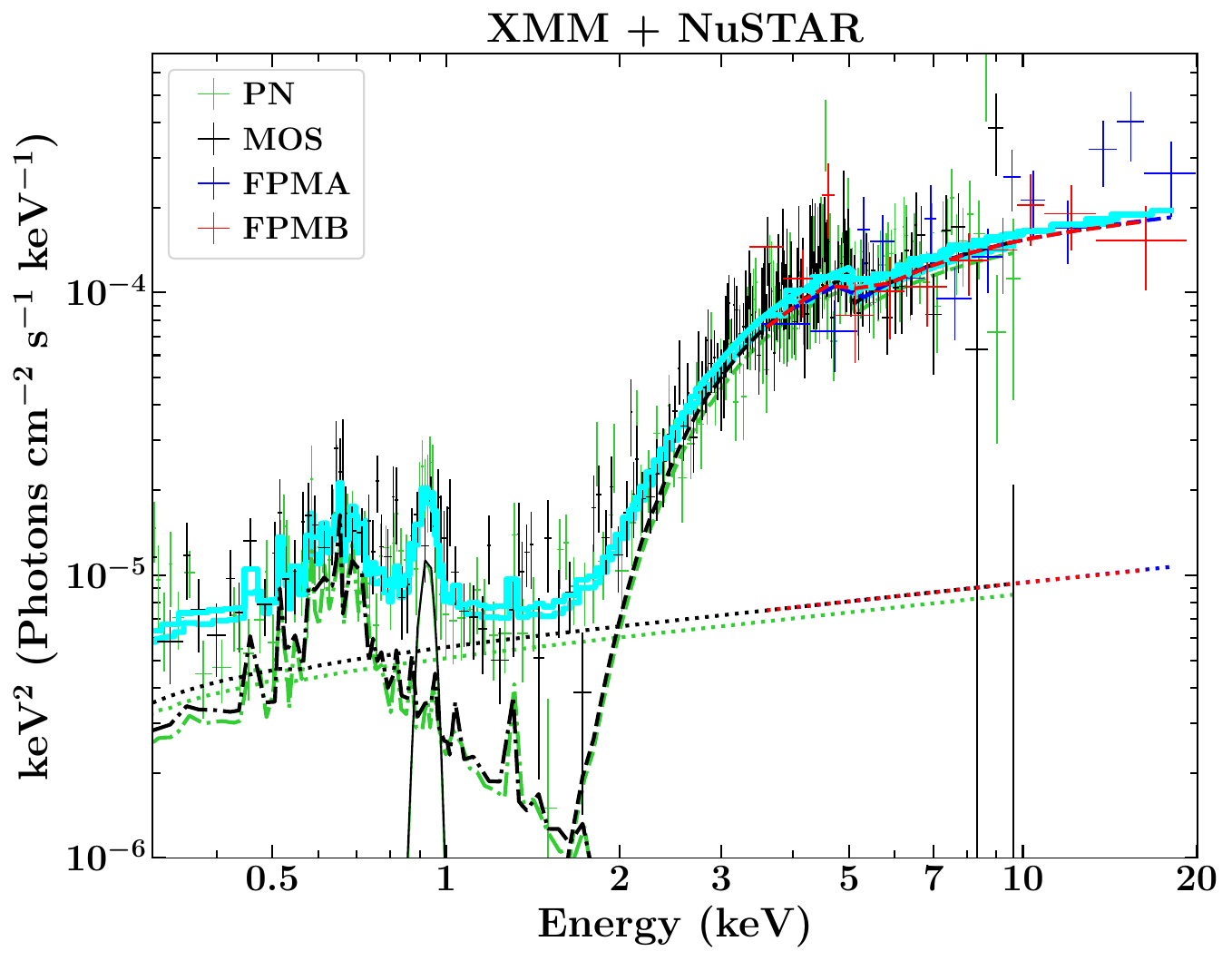}
\caption{Unfolded NuSTAR and Chandra (left) and \xmm~ and NuSTAR (right) spectra of J1128+5831 fitted with an absorbed power law model. The {\tt mekal} and scattered power law components are shown with dash-dot and dotted lines, respectively. The Fe XXII $-$ Fe XXIV L-shell transition is plotted with the solid line and the absorbed power law component is represented with a dashed line. The solid cyan line is the sum of all the model components.}\label{fig:spec}
\end{figure*}

\subsection{An Obscured AGN}
We analyzed the Chandra, \xmm, and NuSTAR observations of J1128+5831 to explore the origin of the X-ray emission. For Chandra and \xmm, we first estimated the source flux based on individual observations, then combined them to improve the signal-to-noise ratio of the spectrum, since no significant flux variability was noticed. The net exposure for the combined EPIC-PN, MOS, and Chandra spectra were 99 ksec, 147 ksec, and 125 ksec, respectively. On the other hand, the NuSTAR data had a total exposure of 60 ksec.

{\it Chandra + NuSTAR}: We fitted a phenomenological model on the joint Chandra and NuSTAR spectrum. The model included a power law ({\tt zpowerlaw} in XSPEC) and absorption due to intervening gas present in the host galaxy of the source ({\tt zphabs}). A small fraction ($\lesssim 5\%$) of the main power law AGN emission can get scattered by the obscuring gas rather than being absorbed by it \citep[e.g.,][]{1997ApJS..113...23T,2018ApJ...854...49M}. This was modeled in XSPEC by adding a second power law component whose photon index was tied to that of the main power law. The fraction of the scattered power law component ($\zeta_{\rm scat}$) was estimated by multiplying it by a constant. Moreover, we also considered a constant to take into account the cross-calibration differences between Chandra and NuSTAR, as well as potential variability between the averaged Chandra observations and the NuSTAR one. This parameter was kept frozen to unity for NuSTAR modules since the flux differences between FPMA and FPMB modules are usually $<$5\% \citep[][]{2017AJ....153....2M}. However, it was allowed to vary for the Chandra dataset. In XSPEC nomenclature, the model configuration is as follows:

\begin{equation}
\begin{aligned}
Model A=&constant1 * phabs*(zphabs*zpowerlaw +\\ 
&constant2*zpowerlaw)
\end{aligned}
\end{equation} 
where $phabs$ was used to model the Galactic absorption and kept fixed during the fit. The fitted model is shown in the left panel of Figure~\ref{fig:spec} and the spectral parameters are provided in Table~\ref{tab:spec_param}. The derived cross-calibration constant of 1.09$^{+0.23}_{-0.18}$ is similar to that found in earlier studies \citep[e.g.,][]{2017AJ....153....2M}. The best-fit reduced $\chi^2$ of the fitted model was derived to be $\chi^2/$degree of freedom (dof) $= 91.2/98 = 0.93$. The best-fit power law photon index is $\Gamma = 1.44^{+0.23}_{-0.23}$, which is similar to that typically found for radio-loud AGNs, including GRSs \citep[e.g.,][]{2018MNRAS.481.4250U}. The intrinsic neutral hydrogen column density was estimated to be $N_{\rm H}=12.02^{+2.20}_{-1.94}\times10^{22}$ cm$^{-2}$. Such a large column density indicates that J1128+5831 hosts an obscured AGN. Furthermore, the scattered fraction was constrained to be $\zeta_{\rm scat}=2.13^{+0.01}_{-0.01}\%$, which is consistent with that reported for obscured AGNs in the literature \citep[e.g.,][]{2017ApJS..233...17R,2018ApJ...854...49M}.

{\it \xmm + NuSTAR}:
Unlike Chandra data, in which the source was detected only above one keV, \xmm~data permitted us to explore the origin of the X-ray emission down to 0.3 keV. The \xmm~spectrum revealed a significant excess emission below 2\,keV, a feature which is typically observed in the X-ray spectra of obscured Type 2 AGN, and is likely to be produced by hot diffuse gas \citep[e.g.,][]{2005A&A...444..119G,2007MNRAS.374.1290G}. In XSPEC, such a thermal feature is usually modeled with {\tt mekal} \citep[][]{1985A&AS...62..197M}. Furthermore, a bump was observed at $\sim$0.9 keV (rest-frame $\sim$1.27 keV), which could be due to Fe XXII $-$ Fe XXIV L-shell transition. To model this feature, we added a Gaussian line whose width was fixed to 30 eV, since we did not find any improvement in the fitting if this parameter was left free to vary. Overall, we modified the model configuration adopted in the joint Chandra and NuSTAR spectral analysis to include these observational features. In XSPEC, the corresponding model setup is as follows:

\begin{equation}
\begin{aligned}
Model B=&constant1 * phabs*(zphabs*zpowerlaw +\\ 
&constant2*zpowerlaw + zgauss + mekal).
\end{aligned}
\end{equation} 

The cross-calibration constant was kept frozen to one for NuSTAR FPMs, and was a free parameter for the EPIC-PN. The best-fitted value was estimated to be 0.91$^{+0.05}_{-0.04}$ which is consistent with that reported in literature \citep[][]{2017AJ....153....2M}. We plot the fitted model in the right panel of Figure~\ref{fig:spec} and report the derived spectral parameters in Table~\ref{tab:spec_param}. The above-mentioned model reproduces the data well ($\chi^2$/dof=612.2/612) and confirms the obscured nature of AGN ($N_{\rm H}=15.58^{+2.46}_{-2.19}\times10^{22}$ cm$^{-2}$). The best-fit power law photon index was marginally consistent to that estimated from the joint Chandra and NuSTAR fitting. However, it is consistent with that usually observed from radio-quiet AGN. Moreover, the fraction of the main AGN emission scattered by the obscuring medium was slightly higher, $\zeta_{\rm scat}=5.72^{+0.02}_{-0.02}\%$, though within the usual range found for GRSs and obscured AGNs \citep[cf.][]{2018MNRAS.481.4250U}.

The application of the {\tt mekal} model revealed the presence of a hot environment ($kT=0.77^{+0.11}_{-0.09}$ keV) with a metallicity which is only $\sim$10\% of the solar value. Such a low relative abundance could indicate the photo-ionization of the hot gas due to quasar emission and has been reported for radio galaxies and obscured AGNs \citep[e.g.,][]{2019ApJ...870...60Z,2024MNRAS.528.1863R,2025MNRAS.536.2025M}.

To compare the X-ray spectral properties of J1128+5831 with that found for powerful radio-loud AGNs, we also fitted an XSPEC model $constant*phabs*(zpow+zphabs*zpow)$, similar to that adopted in previous studies \citep[e.g.,][]{2006MNRAS.370.1893H,2009MNRAS.396.1929H,2014MNRAS.440..269M}. The model was applied to the joint \xmm~and \nustar~spectrum over 0.3$-$20 keV energy range. We obtained the main power law photon index as $2.12^{+0.10}_{-0.09}$ and the intrinsically absorbed power law photon index of $0.86^{+0.03}_{-0.03}$ with the intrinsic neutral hydrogen column density as $7.91^{+0.94}_{-0.82}\times10^{22}$ cm$^{-2}$. However, the fitting was poor ($\chi^2/dof=868/619$), likely due to features present below 2 keV in the \xmm~spectrum. Therefore, we repeated the fit in 2$-$20 keV energy band and obtained a slightly improved fitting result ($\chi^2/dof=345/386$). The best-fitted intrinsic neutral hydrogen column density was $16.54^{+4.19}_{-4.18}\times10^{22}$ cm$^{-2}$ and the main and absorbed power law indices as $4.41^{+2.17}_{-0.51}$ and $1.82^{+0.18}_{-0.17}$, respectively. We found that the entire 2$-$20 keV spectrum can be well explained by a single absorbed power law model, and the power law component with the large photon index was not needed.

The X-ray spectral properties of J1128+5831 are similar to those found for hard X-ray selected GRSs and normal FR~II radio sources \citep[see, e.g.,][]{2006MNRAS.370.1893H,2018MNRAS.481.4250U}. The spectral fitting results suggest J1128+5831 to be an obscured AGN. This finding is consistent with the results reported by \citet[][]{2006MNRAS.370.1893H} who studied a sample of lobe-dominated FR~II sources from the third Cambridge catalog and found narrow-line radio galaxies to be ubiquitously harboring obscured nucleus \citep[see also,][]{2014MNRAS.440..269M}. Moreover, the Fe K-$\alpha$ emission line was not detected in any of the X-ray spectra, which is aligned with the fact that such spectral features are usually weak in off-axis jetted AGNs \citep[][]{2001ApJ...556...35G}.  

The derived photon index of the main power law component is consistent with that usually estimated for radio-quiet AGNs \citep[$\sim1.8$; e.g.,][]{2009ApJ...700L...6R}, thus indicating the observed X-ray emission to be dominated by the disk-corona interaction, similar to that found for hard X-ray selected GRSs \citep[][]{2018MNRAS.481.4250U}. We note that more complex spectral models, e.g., {\tt pexrav} or {\tt myTorus}, could not be fitted due to the faintness of the source, especially above 10 keV, and also due to non-detection of spectral features such as Fe K-$\alpha$ emission line and reflection hump. However, such models become most effective for sources with higher level of obscuration than the one reported here, so we do not expect our results to be significantly affected by the choice of a more simplified model.

\begin{figure*}
\hbox{
\includegraphics[scale=0.4]{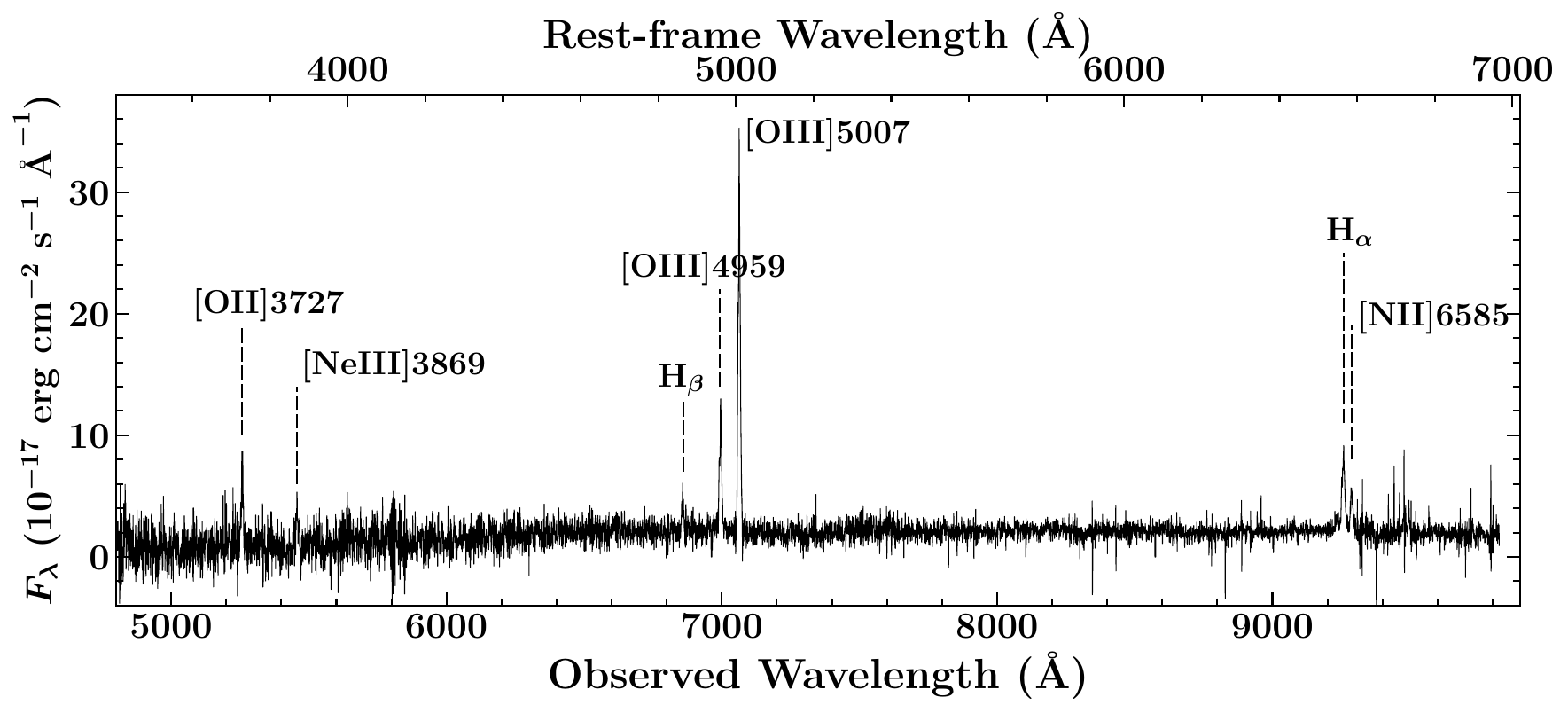}
\includegraphics[scale=0.32]{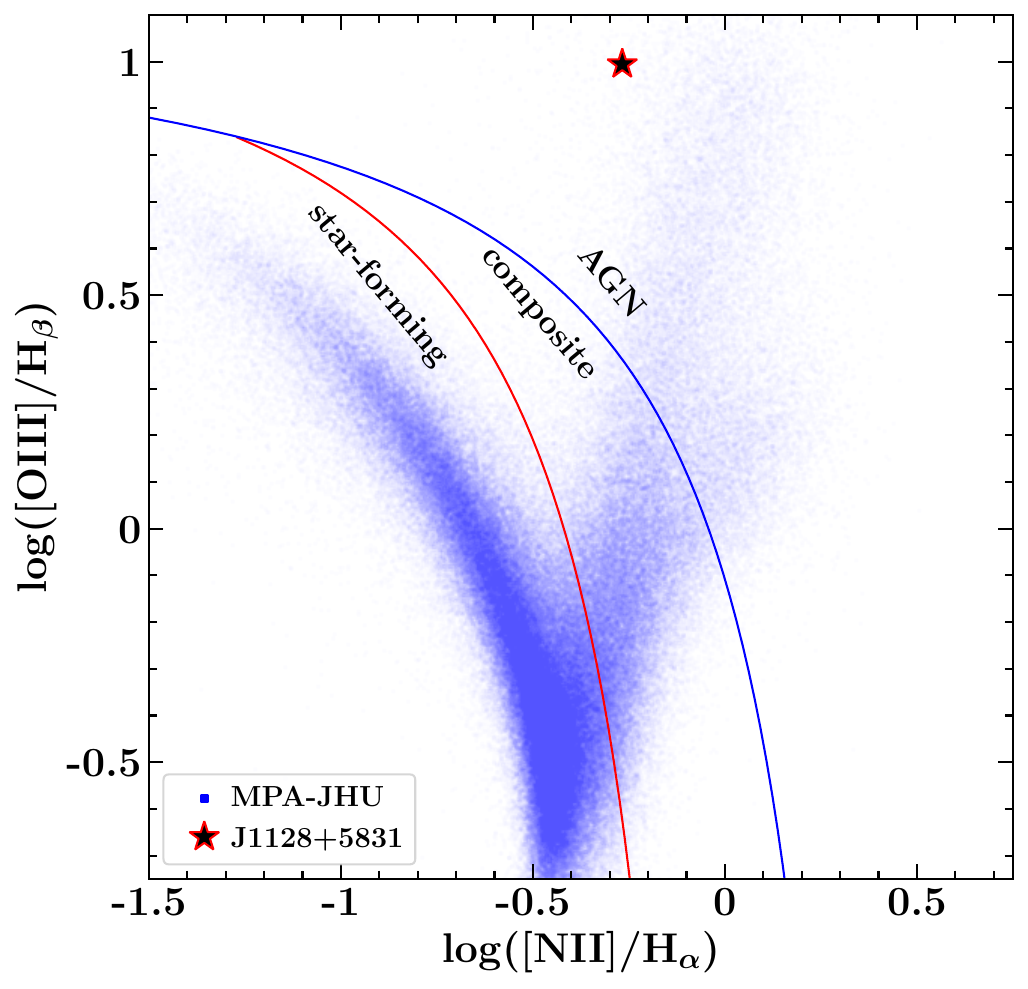}
    }
\caption{The optical spectrum of J1128+5831 is shown in the left panel. The right panel highlights the location of the source in the BPT diagram. The blue dots correspond to SDSS galaxies studied by the group from the Max Planck Institute for Astrophysics, and The Johns Hopkins University \citep[MPA-JHU;][]{2004MNRAS.351.1151B}. The blue and red curves refer to the relations proposed to classify emission-line galaxies \citep[][]{2006MNRAS.372..961K}.}\label{fig:opt}
\end{figure*}

\subsection{Central Engine and Host Galaxy}
One of the hypotheses for Mpc-scale structure in GRSs is the presence of a powerful central engine, i.e., a supermassive black hole and accretion disk, at their centers. The hard X-ray selected GRSs were found to be powered by highly accreting systems, possibly due to rejuvenated AGN activity \citep[see, e.g.,][]{2019MNRAS.489.4049H,2020MNRAS.494..902B}. Therefore, it is crucial to probe the central engine behavior of J1128+5831 and compare it with the general GRS population. The optical spectroscopic observation of J1128+5831 was carried out as part of the DESI survey \citep[][]{2025arXiv250314745D}. We show the optical spectrum in the left panel of Figure~\ref{fig:opt}, which clearly reveals the presence of several prominent narrow emission lines, e.g., \OIIab, \OIIIab, and \NeIII. Furthermore, no broad emission lines were identified, and \halpha~and \hbeta~lines appeared narrow. This observation provides another evidence about the Type 2 nature of J1128+5831. The optical spectrum did not show any signatures of the host galaxy, e.g., Ca H\&K doublet, though the continuum appeared red. 

We fitted the DESI spectrum with the software Bayesian AGN Decomposition Analysis for SDSS Spectra \citep[{\tt BADASS};][]{2024ascl.soft12004S}. The spectrum was corrected for Galactic extinction and brought to the rest-frame using $z=0.4103$. We modeled the continuum with a power law mimicking the AGN emission and a host galaxy template following \citet[][]{2016MNRAS.463.3409V}. The 
emission lines were fitted with a single Gaussian function each, and their signal-to-noise ratios were computed. The uncertainties in the derived parameters were computed following a Monte Carlo bootstrap technique built in {\tt BADASS}. Further details about {\tt BADASS} can be found in \citet[][]{2021MNRAS.500.2871S}.

Since the optical spectrum of J1128+5831 is devoid of broad emission lines, we adopted the procedures described in \citet[][]{2025ApJ...981..101L} to estimate the black hole mass of the source. In particular, the following formula was used \citep[see also,][]{2019MNRAS.487.3404B}:

\begin{equation}\label{eq:baron2019}
  \begin{aligned}
    \log \left(\frac{M_{\rm BH}}{M_{\odot}}\right) = 6.90\, +
    \, 0.54 \times \, \log \left(\frac{\lambda L_{\lambda}}{10^{44}\, {\rm erg \, s^{-1}}} \right) \\
    + 2.06 \, \times \, \log \left(\frac{{\rm FWHM_{\rm H\alpha,br}}}{10^{3} \, {\rm km \, s^{-1}}}\right)
    \end{aligned}
\end{equation}
where $\lambda L_{\lambda}$ is the continuum luminosity at 5100 \AA, and FWHM$_{\rm H\alpha,br}$ is the FWHM of the broad \halpha~component. Since it was not detected, we used the following relation to estimate FWHM$_{\rm H\alpha,br}$ from the ratio of \OIIIb~and narrow \hbeta~line luminosities \citep[][]{2019MNRAS.487.3404B,2025ApJ...981..101L}:

\begin{equation}
  \begin{aligned}\label{eq:o3_nhb}
    \log \left(\frac{\rm FWHM_{\rm H\alpha,br}}{ {\rm 10^3 \, km \, s^{-1}}} \right) = (1.72 \, \pm \, 0.21) \, \times \log \left(\frac{L_{\rm [OIII]}}{L_{\rm H\beta,na}}\right) \\ - (0.62 \pm 0.19)
    \end{aligned}
\end{equation}

From the {\tt BADASS} fitting, we estimated the luminosities of the \OIIIb~and narrow \hbeta~lines as $42.26\pm0.03$ and $41.26\pm0.06$ (\lum, in log-scale), respectively. Inserting these numbers in Equation~\ref{eq:o3_nhb} gives the FWHM of the broad \halpha~component as $12589\pm8864$ km s$^{-1}$. Moreover, since the ultraviolet to optical emitting accretion disk is usually hidden in Type 2 AGNs, the $\lambda L_{\lambda}$ may not be measured directly from the optical spectrum. \citet[][]{2025ApJ...981..101L} proposed the following relation using rest-frame 2$-$10 keV luminosity as a proxy for deriving $\lambda L_{\lambda}$:

\begin{equation}
  \begin{aligned}\label{eq:l5100_xray}
    \log\left(\frac{\lambda L_{\lambda}}{10^{44} \, {\rm erg \, s^{-1}}}\right) = 1.316 \, \times
    \log\left(\frac{L_{\rm 2-10~keV}}{10^{43} \, {\rm erg \, s^{-1}}}\right) - 1.378
  \end{aligned}
\end{equation}

Based on the joint \xmm~and NuSTAR spectral fitting, the rest-frame 2$-$10 keV luminosity is $L_{\rm 2-10~keV}\approx (8\pm1)\times10^{43}$ \lum, which gives $\lambda L_{\lambda}=(6.46\pm0.11)\times10^{43}$ \lum~(Equation~\ref{eq:l5100_xray}). Inserting FWHM$_{\rm H\alpha,br}$ and $\lambda L_{\lambda}$ values in Equation~\ref{eq:baron2019}, we derived the central black hole mass to be $\log M_{\rm BH}=9.06\pm0.63$ (in \Msun). Therefore, we conclude that J1128+5813 harbors about a billion solar mass black hole at its center, similar to that found for the GRS population \citep[cf.][]{2012MNRAS.426..851K,2020A&A...642A.153D,2022A&A...660A..59M}. 

The bolometric luminosity ($L_{\rm bol}$) of J1128+5831 derived from \OIIIb~line luminosity using the relation $L_{\rm bol}=3500 L_{\rm [OIII]}$ is $L_{\rm bol}=6.4\times10^{45}$ \lum~\citep[][]{2004ApJ...613..109H}. However, the use of narrow emission line luminosity as a proxy for $L_{\rm bol}$ has been questioned \citep[see, e.g.,][]{2011MNRAS.412L.123P}. Therefore, we estimated $L_{\rm bol}$ from the IR observations taken with Wide-field Infrared Survey Explorer \citep[WISE; cf.][]{2012MNRAS.422..478R,2014MNRAS.440..269M}. In particular, we adopted the $W4$-filter ($\lambda_{\rm eff}\approx22\mu$m) magnitude, and used the following equation to derive the $L_{\rm bol}$ \citep[][]{2014MNRAS.440..269M}:

\begin{equation}
    \log(L_{\rm bol}) = (15.035\pm4.766) + (0.688\pm0.106)\log(\lambda L_\lambda).
\end{equation}
 We obtained $L_{\rm bol}=5.5\times10^{45}$ \lum, which is similar to that calculated using \OIIIb~line luminosity. For $M_{\rm BH}=10^9$ \Msun, the Eddington luminosity is $L_{\rm Edd}=1.3\times10^{47}$ \lum. Therefore, the Eddington ratio for J1128+5831 is $\eta\equiv L_{\rm bol}/L_{\rm Edd}=0.04$, which suggests the accretion process to be radiatively efficient. Indeed, the detection of bright narrow emission lines in the optical spectrum hints at a rapidly accreting central engine illuminating the broad- and narrow-line regions. In other words, J1128+5831 belongs to the high-excitation radio galaxy (HERG) class of radio sources \citep[e.g.,][]{2012MNRAS.421.1569B}.

Since the optical spectrum of J1128+5831 consists of bright narrow emission lines of the \hbeta~and \halpha~regions, we checked the location of the source in the BPT diagram \citep[][]{1981PASP...93....5B} to determine the possible contribution of star-formation activities to the observed emission lines. In the right panel of Figure~\ref{fig:opt}, we show the BPT diagram, which provides an unambiguous confirmation of the AGN-driven origin of the observed emission lines.

\begin{figure}
\includegraphics[width=\linewidth]{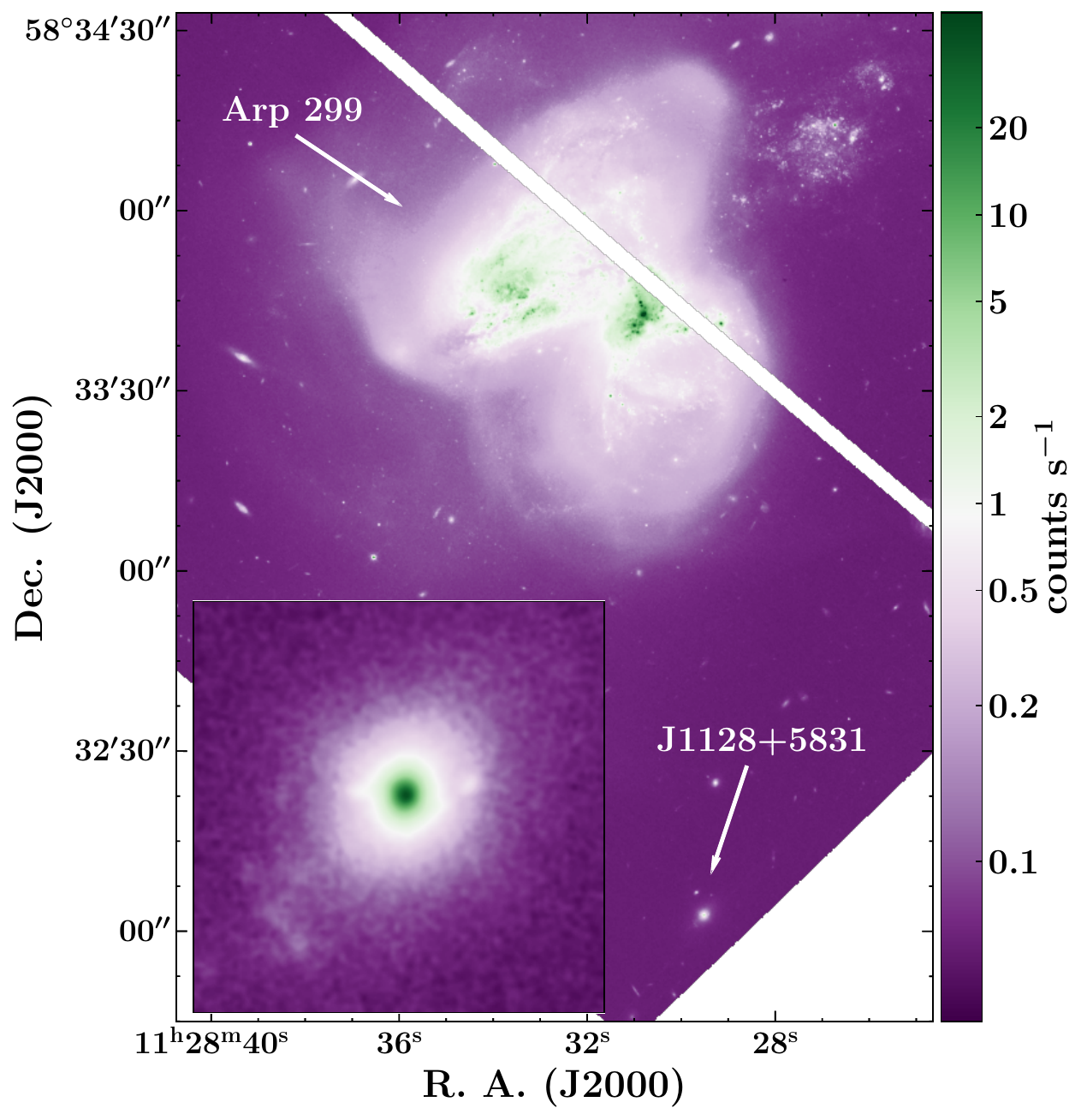}
\caption{The HST image of Arp 299 region taken with the F814W filter. The location of J1128+5831 is highlighted. The inset shows the 5$\times$5 arcsec$^2$ image centered at the host galaxy of the GRS. North is up and east is to the left.}\label{fig:hst}
\end{figure}

Figure~\ref{fig:hst} shows the F814W filter image of Arp 299 region taken with HST in which J1128+5831 is located close to the chip edge. From a quick look, the host appears like an elliptical galaxy which is usually the case for GRSs \citep[][]{2020A&A...635A...5D}. Interestingly, several filamentary structures can be seen towards the southeast of the host galaxy and also along the east and west directions. Such features typically represent the remnants of galaxy mergers \citep[e.g.,][]{2008ApJ...677..846B}. Deeper photometric and spectroscopic observations may reveal the connection of the filamentary structures with the host galaxy of of J1128+5831.

\subsection{Jet Power versus Nuclear Luminosity}
The hard X-ray selected GRSs systematically exhibit a higher nuclear luminosity compared to their time-averaged kinetic jet power ($P_{\rm kin}$) measured from their extended radio luminosity \citep[][]{2018MNRAS.481.4250U}. This observation has been interpreted based on restarted AGN activity, i.e., the objects are currently rapidly accreting with large nuclear luminosity with respect to the past activity that produced large-scale radio jets \citep[e.g.,][]{2019ApJ...875...88B,2020MNRAS.494..902B}. Alternatively, a lower $P_{\rm kin}$ could also be due to significant expansion and radiative losses. We tested this hypothesis for J1128+5831. The $P_{\rm kin}$ can be estimated using the following formula \citep[][]{2018MNRAS.475.2768H,2023JApA...44...13D}:

\begin{equation}
    P_{\rm kin} = 3.3\times10^{44} L_{\rm 151}~{\rm erg~s^{-1}}
\end{equation}
where $L_{\rm 151}$ is the 151 MHz luminosity in units of  10$^{27}$ W Hz$^{-1}$. For J1128+5831, $L_{\rm 151~MHz}\sim1$ (Equation~\ref{eq:power}). This gives $P_{\rm kin}=3.3\times10^{44}$ \lum. Furthermore, the X-ray bolometric correction leads to $L_{\rm bol,X}=5.1\times10^{45}$ \lum~\citep[][]{2025ApJ...981..101L}. A comparison of $P_{\rm kin}$ with $L_{\rm bol,X}$ revealed the latter to be larger, which is similar to that found for other GRSs \citep[][]{2018MNRAS.481.4250U}. 

At radio frequencies the most convincing examples of restarted AGN activity are the double-double radio sources with two or more distinct pairs of lobes \citep[e.g.][]{2009BASI...37...63S}. \cite{2020MNRAS.494..902B} found only three of their fifteen sources to exhibit a clear double-double structure. Eight of their sources were found to exhibit a peaked-spectrum core \citep{2019ApJ...875...88B}. A peaked-spectrum source is suggestive of a young double but not conclusive as these may also exhibit a core-jet or unresolved structure \citep[e.g.][]{2006A&A...454..729X,2009MNRAS.397.2030H}.  J1128+5831 has a very weak radio core with no clear evidence of recurrent AGN activity in the large-scale structure.

If the filamentary features observed in the HST image are associated with the host galaxy of J1128+5831, and originated due to galaxy merger in the past, it may indicate a fresh supply of gas. However, in a recent study of 111 double-double radio galaxies identified from LoTSS DR2, only six showed signs of interactions and mergers in the form of tidal tails and close companion galaxies from the existing data \citep{2025A&A...696A..97D}. Further studies are required to understand whether the high X-ray luminosity may be due to a fresh supply of gas and whether this may trigger AGN activity.

\section{Summary}\label{sec5}
We have conducted a multiwavelength study to explore the physical properties of J1128+5831 ($z=0.4103$), the only GRS serendipitously detected with NuSTAR. We summarize our main results below.

\begin{enumerate}
    \item Located just $\sim$1.5 arcminutes away from Arp 299, J1128+5831 hosts a radio source with an end-to-end projected length of 864 kpc. The source has a weak core dominance ($C_{\rm D}=-2.4$) and an overall steep radio spectrum ($\alpha=-0.86$). These results indicate J1128+5831 to be viewed at a large angle from the jet axis, similar to Type 2 AGNs.
    \item The X-ray spectral analysis of the joint Chandra and NuSTAR, and \xmm~and NuSTAR datasets revealed the presence of an obscured AGN with neutral hydrogen column density exceeding 10$^{23}$ cm$^{-2}$. Moreover, the application of the {\tt mekal} model suggested the existence of a hot ($kT=0.77^{+0.11}_{-0.09}$ keV) gas with sub-solar metallicity (relative abundance = 0.12$^{+0.12}_{-0.07}$), possibly due to photo-ionization of the hot gas by AGN emission.
    \item The optical spectrum of J1128+5831 consists of bright narrow emission lines devoid of broad components, thus providing further evidence for the source to be a Type 2 AGN. It harbors a massive black hole ($M_{\rm BH}=10^9$ \Msun) and a radiatively efficient accretion system with Eddington ratio of 0.04. Therefore, J1128+5831 can be considered to be a HERG.
    \item In the BPT diagram, J1128+5831 lies in a region populated by AGNs, thus hinting that the photo-ionization of the narrow line region clouds produces the observed emission lines.
    \item The broadband properties of J1128+5831 are consistent with those observed for the general GRS population \citep[e.g.,][]{2018MNRAS.481.4250U,2023JApA...44...13D}.
\end{enumerate}

\acknowledgements
We thank the journal referee for constructive criticism. 
This work has made use of data from the NuSTAR mission, a project led by the California Institute of Technology, managed by the Jet Propulsion Laboratory, and funded by the National Aeronautics and Space Administration. This research has made use of the NuSTAR Data Analysis Software (NuSTARDAS) jointly developed by the ASI Science Data Center (ASDC, Italy) and the California Institute of Technology (USA). This work is based on observations obtained with XMM-Newton, an ESA science mission with instruments and contributions directly funded by ESA Member States and NASA. The scientific results reported in this article are based on data obtained from the Chandra Data Archive. This research has made use of software provided by the Chandra X-ray Center (CXC) in the application packages CIAO, ChIPS, and Sherpa. This research made use of the cross-match service provided by CDS, Strasbourg.

LOFAR data products were provided by the LOFAR Surveys Key Science project (LSKSP; https://lofar-surveys.org/) and were derived from observations with the International LOFAR Telescope (ILT). LOFAR is the Low Frequency Array designed and constructed by ASTRON. It has observing, data processing, and data storage facilities in several countries, which are owned by various parties (each with their own funding sources), and which are collectively operated by the ILT foundation under a joint scientific policy. The efforts of the LSKSP have benefited from funding from the European Research Council, NOVA, NWO, CNRS-INSU, the SURF Co-operative, the UK Science and Technology Funding Council and the J\"{u}lich Supercomputing Centre. 

We acknowledge the use of Hubble Advanced Products available at Mikulski Archive for Space Telescopes. This research has made use of NASA's Astrophysics Data System Bibliographic Services. This research has made use of data obtained through the High Energy Astrophysics Science Archive Research Center Online Service, provided by the NASA/Goddard Space Flight Center.

The HST data is available at MAST: \dataset[doi: 10.17909/avwb-jy09]{\doi{10.17909/avwb-jy09}}.
This paper employs a list of Chandra datasets, obtained by the Chandra X-ray Observatory, contained in~\dataset[DOI: 10.25574/cdc.532]{https://doi.org/10.25574/cdc.532}.

\facilities{\xmm, NuSTAR, Chandra, HST}
\software{CIAO (v.4.17), SAS (v.21.0.0), XSPEC \citep[v 12.10.1;][]{1996ASPC..101...17A}, Astropy \citep[][]{2013A&A...558A..33A,2018AJ....156..123A}}

\bibliographystyle{aasjournal}
\bibliography{Master}
\end{document}